\documentstyle[preprint,aps]{revtex}

\begin{document}

\draft
\preprint{LBL-34409}

\title{\nopagebreak Minijet-associated Dilepton Production in
Ultrarelativistic Heavy Ion Collisions \thanks{\baselineskip=12pt
This work was supported by the Director, Office of Energy
Research, Division of Nuclear Physics of the Office of High
Energy and Nuclear Physics of the U.S. Department of Energy
under Contract No. DE-AC03-76SF00098.}}

\author{K. J. Eskola$^{1,2}$ and Xin-Nian Wang$^1$}

\address{$^1$Nuclear Science Division, Mailstop 70A-3307,
        Lawrence Berkeley Laboratory\\
University of California, Berkeley, California 94720.\\
$^2$Laboratory of High Energy Physics, P.O. Box 9,
        SF-00014 University of Helsinki, Finland.\cite{address}}

\date{Oct. 15, 1993}

\maketitle

\begin{abstract}

        Dilepton production associated with minijets is
calculated in ultrarelativistic heavy ion collisions using
the first order approximation of the dilepton fragmentation
functions of quarks and gluons. The full QCD evolution of
the fragmentation functions is also studied. We find
that the dilepton pairs from the fragmentation of minijets
are comparable to direct Drell-Yan at $\sqrt{s}=200$ AGeV for
small dilepton invariant mass $M\sim$ 1-2 GeV/c$^2$ while
dominant over a large range of mass at $\sqrt{s}=6400$ AGeV.

\end{abstract}
\narrowtext

\section{INTRODUCTION}

        In the search for a quark gluon plasma (QGP) in
ultrarelativistic heavy ion collisions, electromagnetic
signals are considered good probes of the dense matter \cite{RUUS}.
Because of the large electromagnetic mean free path, leptons
and photons produced by interacting (anti)quarks inside QGP can
easily escape the hot and dense matter and carry the information
of the system to the detector. Recent developments in parton
transport phenomenology indicate that dileptons and photons
could also reveal the dynamics of the early evolution of the
a dense parton system \cite{LMCL,KGJK,SHXL}. However,
like for all the proposed
QGP signals, background must be understood and subtracted
in order to distinguish the true features of a QGP. In
general, there are two kinds of background sources for
the thermal electromagnetic signals. One comes from the evolution of
the hadronic phase and the decays of the produced hadrons.
The other one is due to the initial parton scatterings at
the very earliest stage of the heavy ion collisions.
For dilepton production, the latter one is usually
referred to as Drell-Yan (DY) processes \cite{DY}.

        In the lowest order, ${\cal O}(\alpha^2)$, the DY
processes are simply  quark-antiquark annihilations. First
order contributions ${\cal O}(\alpha^2\alpha_s)$ in
perturbative QCD (pQCD), originating from initial
state radiation and virtual corrections, give
rise to about the same amount of dilepton production as in
the lowest order, which is often characterized by a so-called
``$K$-factor'' of  about 2 \cite{DYK1,FKWS}.
These corrections are also responsible for large
$p_T$ tails of the dilepton transverse momentum spectrum.
At  small $p_T$, summation over the initial state soft gluon
radiations generates a  Sudakov form factor regularizing
the perturbative low $p_T$ production  \cite{DYpT}.
By now, there also exist  matrix element  calculations
of the second order pQCD contributions,
${\cal O}(\alpha^2\alpha_s^2)$,
to the $K$-factor \cite{DYK2}.

In this paper, we will investigate dilepton production
associated with minijet final state radiation in
heavy  ion collisions at collider energies.
It is expected that at energies
$\sqrt{s} \,{\buildrel > \over {_\sim}}\,200$ AGeV, minijets
[(anti)quarks and gluons with $p_T\sim$  few GeV/c] are
produced abundantly via
multiple semihard scatterings. These  minijets
have important contributions to  particle production,
transverse energy and overall evolution of the formed
quark-gluon system \cite{KAJA}-\cite{RANF}.
Therefore, it would be interesting to study
dilepton bremsstrahlung from the initially
produced minijets. Especially, it
is important to know whether the dileptons associated
with  minijets could compete with the lower order DY processes
at  midrapidity and small invariant dilepton masses $M\sim 1-3$
GeV/c$^2$, where a window for observing the thermal
dileptons is expected \cite{RUUS}.

Rather than strictly applying the almost complete
${\cal O}(\alpha^2\alpha_s^2)$ results from the matrix element
calculations \cite{DYK2}, we take a different and
simpler approach by calculating the dilepton fragmentation
functions of the final state minijets.
Unlike in the real photon fragmentation functions
\cite{OWEN,REYA},  the relatively large invariant masses
$M\gg\Lambda$ of the dileptons fix the lower limit of the
momentum scale of the QCD radiation processes. This makes
the problem calculable in pQCD. In the leading logarithm
approximation and in  an axial gauge \cite{FIED},
the dilepton fragmentation functions can be calculated up
to all orders in pQCD. Using the obtained fragmentation
functions to convolute with minijet cross sections, we then
compute the contribution to the dilepton production from
the final state radiation of minijets. We will
show how the associated production of $M\sim 1-2$ GeV/c$^2$
dileptons from minijets with $p_T\ge2$ GeV/c is
comparable to the first order DY results at
BNL Relativistic Heavy Ion Collider (RHIC) energies
but becomes dominant at CERN
Large Hadron Collider (LHC) energies, even up to masses
$M\sim 5-10$ GeV/c$^2$.  We will also study the effects of
nuclear modifications of the parton distributions,
especially nuclear shadowing, to the dilepton rates.

The remainder of the paper is organized as follows.
In the next Section, we will calculate the dilepton
fragmentation
functions in the  framework of pQCD. The connection and the
difference between real photon fragmentation functions are
discussed. We will derive both the first order result
and the one with QCD evolution, including corrections to all
orders in pQCD. In Sec. III, the dilepton fragmentation
functions are convoluted with hard and semihard parton
scattering cross sections to calculate the minijet-associated
dilepton production in heavy ion collisions at both
RHIC and LHC energies.
Finally, a summary with some discussions on the
implications to the dilepton production from  a QGP
is given in Sec. IV.

\section{DILEPTON FRAGMENTATION FUNCTIONS}

        In this Section, we review the dilepton fragmentation
functions of quark and gluon jets. We will work in an axial
gauge so that interference terms in the final state radiation
disappear in the leading logarithm approximation \cite{FIED}.

\subsection{Lowest order in pQCD}

        Let us define $z$ as the fractional light-cone
momentum and $q^2$ as the virtuality of an off-shell parton
as illustrated in Fig.~\ref{fig1}(a). The differential cross section
for a quark $q_i$ produced in a hard process with
momentum scale $Q$ to emit a dilepton with invariant
mass $M$ is,
\begin{equation}
        \frac{1}{\sigma_0}\frac{d\sigma}{dz\,dz_{\ell}\,dq^2\,dM^2}
        =e_i^2\frac{\alpha}{2\pi q^2}P_{q\rightarrow\gamma q}(z)
        \frac{\alpha}{2\pi M^2}P_{\gamma\rightarrow\ell^+\ell^-}(z_{\ell}),
                \label{eq:dq01}
\end{equation}
where $e_i$ is the fractional charge of the quark $q_i$,
$\sigma_0$ is the total cross section of the hard process,
and
\begin{eqnarray}
        P_{q\rightarrow\gamma q}(z)&=&\frac{1}{z}[1+(1-z)^2],\\
        P_{\gamma\rightarrow\ell^+\ell^-}(z)&=&z^2+(1-z)^2,
\end{eqnarray}
are the splitting functions for $q\rightarrow\gamma q$ and
$\gamma\rightarrow\ell^+\ell^-$ which are similar to those
of $q\rightarrow g q$ and $g\rightarrow q\bar{q}$,
respectively, except for the color factors. Integrating over
the virtuality $q^2$ of the intermediate quark and the
fractional momentum $z_{\ell}$ of one of the leptons, one
has the QED dilepton fragmentation function of a quark,
\begin{eqnarray}
        D^{(0)}_{{\rm DL}/q_i}(z,M^2,Q^2)&\equiv&
                \int_{M^2}^{Q^2}dq^2\int_0^1dz_{\ell}
                \frac{1}{\sigma_0}\frac{d\sigma}{dz\,dz_{\ell}\,dq^2\,dM^2}
                        \nonumber \\
        &=&e_i^2\left(\frac{\alpha}{2\pi}\right)^2\frac{2}{3M^2}
                \ln\left(\frac{Q^2}{M^2}\right)\frac{1}{z}[1+(1-z)^2].
                        \label{eq:dq02}
\end{eqnarray}
One can see that $D^{(0)}_{{\rm DL}/q_i}(z,M^2,Q^2)$
is similar to a virtual photon fragmentation function,
except for a factor due to the extra QED coupling and the
integration over the relative phase space of the leptons,
\begin{equation}
        D^{(0)}_{{\rm DL}/q_i}(z,M^2,Q^2)=
        \frac{\alpha}{2\pi}\frac{2}{3M^2}
        D^{(0)}_{\gamma^*/q_i}(z,M^2,Q^2).
\end{equation}

For a real photon fragmentation function, the lower limit
for the integration over $q^2$ in Eq.~\ref{eq:dq02} is
in principle given by the quark mass. For massless
quarks, the infrared divergence in the lowest order
has to be regulated by some cutoff of the hadronic scale.
In the absence of a large mass scale, the
QCD corrections to real photon fragmentation function
have also to be regulated by some cutoff. The physics
below the cutoff becomes nonperturbative.
One has to introduce
some initial conditions for the real photon fragmentation
functions, either given by experimental data
or by some model-dependent assumptions. The problem of
dilepton production is different because the fixed
invariant mass $M$ provides a natural cutoff
below which kinematic restrictions will terminate
the processes. The QCD processes above this cutoff
are in principle calculable to all orders.

        Since gluons are not directly coupled to
photons and leptons, the dilepton fragmentation function
of a gluon in the lowest order is,
\begin{equation}
        D^{(0)}_{{\rm DL}/g}(z,M^2,Q^2)=0.
\end{equation}

        For later convenience, we define
\begin{equation}
        \kappa(M,Q)\equiv\ln\left[\frac{\ln(Q^2/\Lambda^2)}
                {\ln(M^2/\Lambda^2)}\right], \label{eq:kappa}
\end{equation}
and
\begin{equation}
        Q_i^{(0)}(z)\equiv e_i^2\frac{1}{z}[1+(1-z)^2], \label{eq:qz0}
\end{equation}
so that we can rewrite Eq.~\ref{eq:dq02} as
\begin{equation}
        D^{(0)}_{{\rm DL}/q_i}(z,M^2,Q^2)=
\left(\frac{\alpha}{2\pi}\right)^2\frac{2}{3M^2}
\ln(\frac{M^2}{\Lambda^2})(e^{\kappa}-1)Q_i^{(0)}(z).
                \label{eq:dq03}
\end{equation}

\subsection{First order contributions}

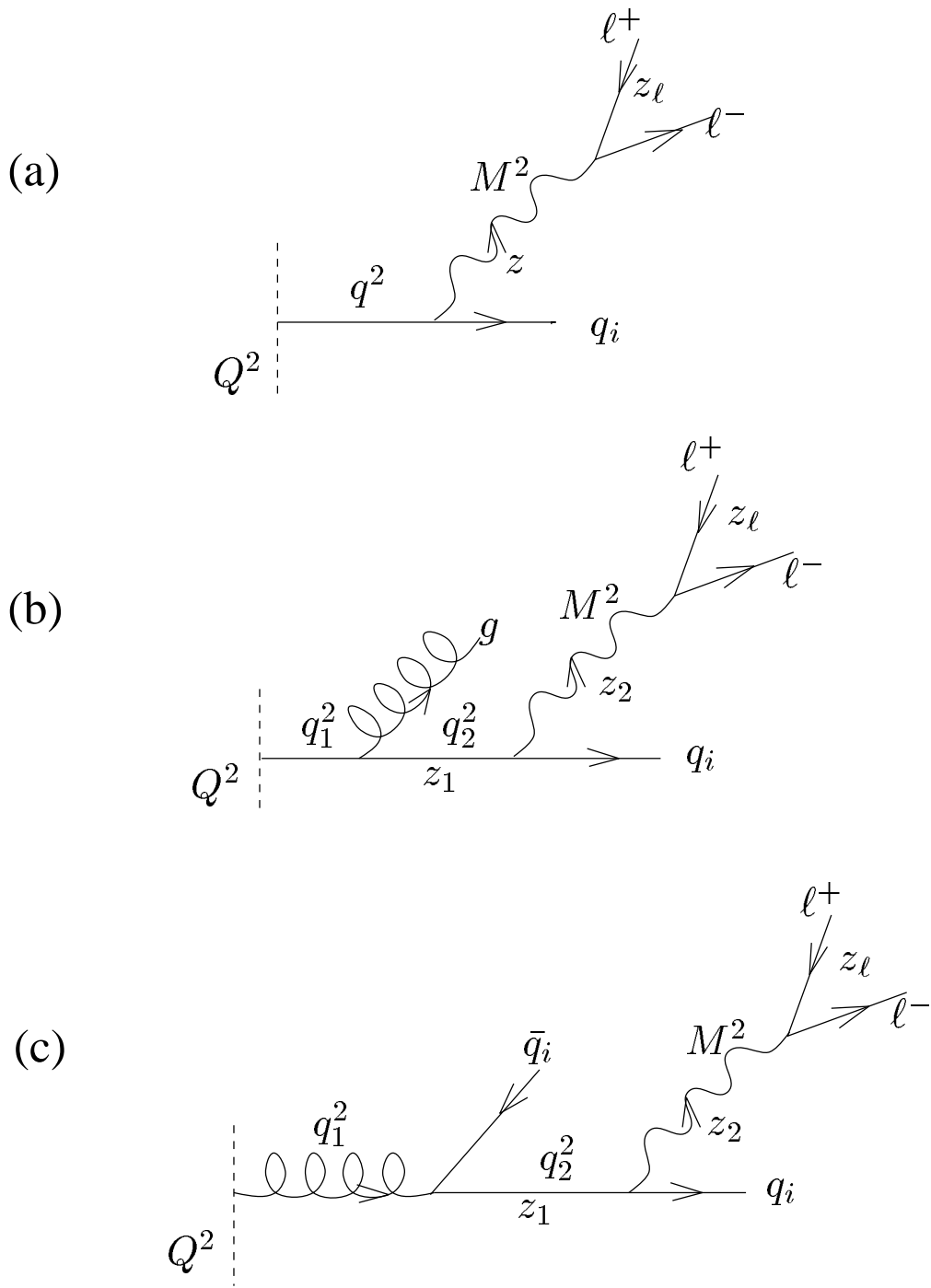
\begin{figure}
\caption{Illustration of the diagrams of (a) the lowest order,
        (b) the first order contributions in pQCD to the dilepton
        fragmentation functions of quarks and (c) gluons. The
        dashed lines present the associated hard processes
        with momentum scale $Q$.}
\label{fig1}
\end{figure}

        The first order contribution in pQCD to the
dilepton fragmentation function of a quark comes
from a gluon bremsstrahlung before the virtual
photon production as shown in Fig.~\ref{fig1}(b). Remember
now that $z=z_1\,z_2$ is the fraction of the momentum,
carried by the dilepton, of the initial quark before
the gluon radiation. Defining the convolution of two
functions as
\begin{equation}
        A\otimes B(z)\equiv\int_z^1\frac{dz_1}{z_1}A(z_1)B(z/z_1),
\end{equation}
it is straightforward to write down the first order
dilepton fragmentation function,
\begin{eqnarray}
        D^{(1)}_{{\rm DL}/q_i}(z,M^2,Q^2)&=&
        \int_{M^2}^{Q^2}dq^2_1 \frac{\alpha_s(q^2_1)}{2\pi q^2_1}
        \int_z^1\frac{dz_1}{z_1}P_{q\rightarrow qg}(z_1)
        D^{(0)}_{{\rm DL}/q_i}(\frac{z}{z_1},M^2,Q^2) \nonumber \\
        &=&\left(\frac{\alpha}{2\pi}\right)^2\frac{2}{3M^2}
        \ln(\frac{M^2}{\Lambda^2})\frac{2}{\beta_0}
        (e^{\kappa}-1-\kappa)P_{q\rightarrow qg}\otimes Q_i^{(0)}(z),
                \label{eq:dq11}
\end{eqnarray}
where,
\begin{equation}
        \alpha_s(q^2)=\frac{4\pi}{\beta_0\ln(q^2/\Lambda^2)},
                        \;\; \beta_0=11-2n_f/3,\label{eq:alphas}
\end{equation}
is the running strong coupling constant with $n_f$ quark
flavors. The splitting function for $q\rightarrow q g$ in QCD is
\begin{equation}
        P_{q\rightarrow qg}(z)=\frac{4}{3}\left[\frac{1+z^2}{1-z}\right]_+.
\end{equation}
The ``+ function'' here is introduced to include the virtual
corrections to cancel the singularity from the soft gluon
emission and to guarantee momentum conservation \cite{FIED,AP}.
Other splitting functions we will use in the following are
the standard ones \cite{AP},
\begin{eqnarray}
        P_{q\rightarrow g q}(z)&=&P_{q\rightarrow qg}(1-z),\\
        P_{g\rightarrow q\bar{q}}(z)&=& \frac{1}{2}[z^2+(1-z)^2],\\
        P_{g\rightarrow gg}(z) &=&6\left[\frac{z}{(1-z)_+}
                        +\frac{1-z}{z} + z(1-z)+(\frac{11}{12}
                        -\frac{2n_f}{36})\delta(1-z)\right].
\end{eqnarray}

        The convolution in Eq.~\ref{eq:dq11} can be easily done
and it gives,
\begin{eqnarray}
        Q_i^{(1)}(z)&\equiv&P_{q\rightarrow qg}\otimes Q_i^{(0)}(z)
                        \nonumber \\
        &=&\frac{4e_i^2}{3z}\left\{ 2[1+(1-z)^2]\ln(1-z)\right.
        + \left. (2-z)z\ln\,z
        +z(2-\frac{1}{2}z)\right\}. \label{eq:qz1}
\end{eqnarray}

        To the first order in pQCD, the dilepton fragmentation
function of a gluon is not zero anymore. From the diagram
in Fig.~\ref{fig1}(c), we have
\begin{eqnarray}
        D^{(1)}_{{\rm DL}/g}(z,M^2,Q^2)&=&
        \sum_{i=1}^{2n_f}\int_{M^2}^{Q^2}dq^2_1
        \frac{\alpha_s(q^2_1)}{2\pi q^2_1}
        \int_z^1\frac{dz_1}{z_1}P_{g\rightarrow q\bar{q}}(z_1)
        D^{(0)}_{{\rm DL}/q_i}(\frac{z}{z_1},M^2,Q^2) \nonumber \\
        &=&\left(\frac{\alpha}{2\pi}\right)^2\frac{2}{3M^2}
        \ln(\frac{M^2}{\Lambda^2})\frac{2}{\beta_0}(e^{\kappa}-1-\kappa)
        \sum_{i=1}^{2n_f}P_{g\rightarrow q\bar{q}}\otimes Q_i^{(0)}(z),
                \label{eq:dg11}
\end{eqnarray}
and the convolution in $z$ can also be calculated explicitly
in this case, giving
\begin{eqnarray}
        G^{(1)}(z)&\equiv&\sum_{i=1}^{2n_f}
        P_{g\rightarrow q\bar{q}}\otimes Q_i^{(0)}(z) \nonumber \\
        &=&\sum_{i=1}^{2n_f}e_i^2\frac{1}{2z}
        [\frac{4}{3}(1-z^3)+z(1-z)+2(1+z)z\ln\,z]. \label{eq:gz1}
\end{eqnarray}

        As we will see below numerically, radiative corrections to
any order will soften the QED fragmentation function of quarks
and increase the fragmentation function of gluons. Because
of the leading logarithm behavior of the radiations,
the dependence of QCD corrections on the strong coupling
constant is cancelled out so that they might become
important to all orders. From
Eqs.~\ref{eq:dq03} and \ref{eq:dq11}, we can see that
the relative importance of the first order QCD
correction to the QED fragmentation
function is controlled by a $Q$-dependent factor,
\begin{equation}
        C\sim 1-\frac{\kappa}{e^{\kappa}-1},
\end{equation}
where $\kappa$ is defined in Eq.~\ref{eq:kappa}. For values
of $Q^2$ not too large relative to $M^2$, $\kappa$ is very small
so that higher order corrections can be neglected. Only for
extremely large values of $Q^2$ and thus $\kappa$, $C$
becomes comparable to 1. Then one has to include
corrections to all orders.  For our consideration here, $Q^2$
is in the order of $p_T^2$ of the minijets. Thus, as we
will show in the next Section, for most of the minijet production with
$p_T\sim$ 2 GeV/c, first order calculation of the
dilepton fragmentation functions should be sufficient.

\subsection{Full QCD evolution}

        Following the same steps as we have calculated the
first order corrections to the dilepton fragmentation functions,
we can calculate the higher order contributions. Here we neglect
the further splitting of the radiated soft gluons and quarks,
and only consider those diagrams with a simple ladder structure
in leading logarithm approximation. Therefore, the radiated
soft gluons and quarks are always on mass-shell. The general
form of the contributions with $n$ radiations before the
dilepton production can be derived as
\begin{eqnarray}
        D^{(n)}_{{\rm DL}/q_i}(z,M^2,Q^2)&=&
        \left(\frac{\alpha}{2\pi}\right)^2\frac{2}{3M^2}
        \ln\left(\frac{M^2}{\Lambda^2}\right)\nonumber \\
        &\times&\left(\frac{2}{\beta_0}\right)^n
        \left[e^{\kappa}-(1+\kappa+\cdots+\frac{\kappa^n}{n!})
                \right] Q_i^{(n)}(z), \label{eq:qn}\\
        D^{(n)}_{{\rm DL}/g}(z,M^2,Q^2)&=&
        \left(\frac{\alpha}{2\pi}\right)^2\frac{2}{3M^2}
        \ln\left(\frac{M^2}{\Lambda^2}\right)\nonumber \\
        &\times&\left(\frac{2}{\beta_0}\right)^n
        \left[e^{\kappa}-(1+\kappa+\cdots+\frac{\kappa^n}{n!})
                \right] G^{(n)}(z), \label{eq:gn}
\end{eqnarray}
where $Q_i^{(n)}(z)$ and $G^{(n)}(z)$ can be calculated
iteratively from the lower order results via
\begin{eqnarray}
        Q^{(n)}_i(z)&=&P_{q\rightarrow qg}\otimes Q^{(n-1)}_i(z)
                +P_{q\rightarrow gq}\otimes G^{(n-1)}(z), \label{eq:qzn}\\
        G^{(n)}(z)&=&\sum_{i=1}^{2n_f} P_{g\rightarrow q\bar{q}}
        \otimes Q^{(n-1)}_i(z) +P_{g\rightarrow gg}\otimes G^{(n-1)}(z).
                                \label{eq:gzn}
\end{eqnarray}
Since we know $Q^{(0)}_i(z)$ (see Eq.~\ref{eq:qz0}) and
$G^{(0)}(z)=0$, we can in principle perform the above
convolutions up to any order as we did for
$Q^{(1)}_i(z)$ (Eq.~\ref{eq:qz1}) and  $G^{(1)}(z)$
(Eq.~\ref{eq:gz1}), and obtain the full QCD dilepton
fragmentation functions as,
\begin{eqnarray}
        D_{{\rm DL}/q_i}(z,M^2,Q^2)&=&
        \sum_{n=0}^{\infty}D^{(n)}_{{\rm DL}/q_i}(z,M^2,Q^2),
                        \label{eq:qfull}\\
        D_{{\rm DL}/g}(z,M^2,Q^2)&=&
        \sum_{n=1}^{\infty}D^{(n)}_{{\rm DL}/g}(z,M^2,Q^2).
                \label{eq:gfull}
\end{eqnarray}
For large values of $n$, evaluating the integration
in the convolution analytically is obviously too cumbersome.
One method to evaluate the full fragmentation
functions is to solve numerically
a set of coupled evolution equations.

        Taking derivatives of $D_{{\rm DL}/q_i,g}(z,M^2,Q^2)$
with respect to $Q^2$ and using
\begin{equation}
        Q^2\frac{d\kappa}{dQ^2}=\frac{1}{\ln(Q^2/\Lambda^2)},
\end{equation}
and the definition of $\alpha_s(Q^2)$ in Eq.~\ref{eq:alphas},
we can derive from Eqs.~\ref{eq:qn}-\ref{eq:gfull} the
following coupled evolution equations,
\begin{eqnarray}
        \frac{dD_{{\rm DL}/q_i}}{d\ln Q^2}(z,M^2,Q^2)
        &=&\left(\frac{\alpha}{2\pi}\right)^2\frac{2}{3M^2}Q_i^{(0)}(z)
        +\frac{\alpha_s(Q^2)}{2\pi}P_{q\rightarrow qg}
                \otimes D_{{\rm DL}/q_i}(z,M^2,Q^2) \nonumber \\
        &+&\frac{\alpha_s(Q^2)}{2\pi}P_{q\rightarrow gq}
                \otimes D_{{\rm DL}/g}(z,M^2,Q^2), \label{eq:evq}\\
        \frac{dD_{{\rm DL}/g}}{d\ln Q^2}(z,M^2,Q^2)
        &=&\frac{\alpha_s(Q^2)}{2\pi}\sum_{i=1}^{2n_f}
        P_{g\rightarrow q\bar{q}}
        \otimes D_{{\rm DL}/q_i}(z,M^2,Q^2)\nonumber \\
        &+&\frac{\alpha_s(Q^2)}{2\pi}P_{g\rightarrow  gg}
                \otimes D_{{\rm DL}/g}(z,M^2,Q^2). \label{eq:evg}
\end{eqnarray}

        These evolution equations are very similar to
those of real photon fragmentation functions \cite{OWEN,REYA}
and the parton distribution functions in a photon \cite{DEWT}.
The only difference is that dilepton (or virtual photon)
fragmentation functions with a given mass $M$ have
a definite initial condition,
\begin{equation}
        \left.D_{{\rm DL}/q_i,g}(z,M^2,Q^2)\right|_{Q^2=M^2}=0,
\end{equation}
together with the boundary condition,
\begin{equation}
        \left.D_{{\rm DL}/q_i,g}(z,M^2,Q^2)\right|_{z=1}=0.
\end{equation}
The boundary condition simply means that the probability
for the dilepton to take the whole fraction of momentum of
the initial quark or gluon is zero after QCD evolution is
taken into account. Since there is always a finite contribution
to $dD_{{\rm DL}/q_i}/dQ^2$ from the QED term in the evolution
equation Eq.~\ref{eq:evq}, one can verify that
$D_{{\rm DL}/q_i}(z,M^2,Q^2)$ must approach zero as
$1/\ln(1-z)$ at $z=1$ in order to satisfy the boundary condition
at all $Q^2$.  For gluons, $D_{{\rm DL}/g}(z,M^2,Q^2)$
must go to zero faster than $1/\ln(1-z)$.

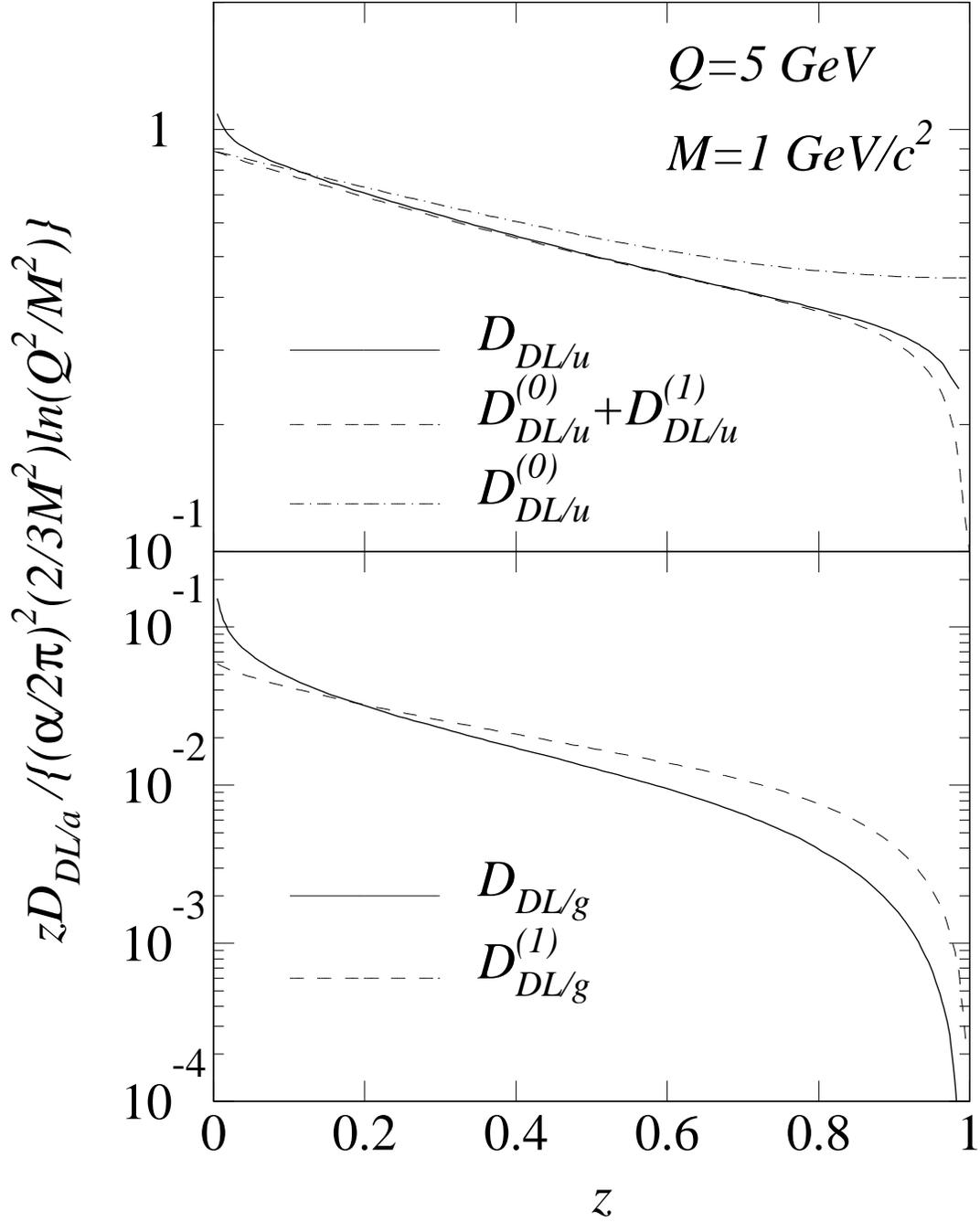
\begin{figure}
\caption{The QCD-evolved (solid), the lowest order (dot-dashed)
        and the first order (dashed) approximations of dilepton
        fragmentation functions  $zD_{{\rm DL}/a}(z,M^2,Q^2)$ of
        a $u$-quark and a gluon, for $M=1$ GeV/c$^2$ and $Q=5$ GeV.
        A factor $(\alpha/2\pi)^2(2/3M^2)\ln(Q^2/M^2)$ is divided out.}
\label{fig2}
\end{figure}

        The above evolution equations with the initial
and boundary conditions can be solved numerically.
The scale $M^2$ now sets the starting
point of the evolution. We show the QCD-evolved
dilepton fragmentation functions
$zD_{{\rm DL}/q_i,g}(z,M^2,Q^2)$ scaled by a
common factor $(\alpha/2\pi)^2 (2/3M^2) \ln(Q^2/M^2)$
in Fig.~\ref{fig2} for $M=1$ GeV/c$^2$ and $Q=$ 5 GeV.
Together, we also show the analytical results to the
lowest and first order. It is clear that
both the first order corrections and the full QCD evolution
soften the fragmentation functions. The overall QCD
corrections to the QED (or lowest order in pQCD) result
are about  10\%, except near $z=0$ and 1.  Since a gluon does
not have dilepton production to the zeroth order in pQCD,
the dilepton fragmentation function of a gluon is
one order of magnitude smaller than a quark.
Because there are logarithmic divergences
at $z=0$ and 1 for each order correction to the dilepton
fragmentation functions, as can be seen
in Eqs.~\ref{eq:qz1} and \ref{eq:gz1}, every order becomes important
so that one has to sum them together to get the
full QCD result. This is why the full QCD-evolved
fragmentation functions in Fig.~\ref{fig2} differ considerably
from the first order results near $z=0$ and 1. To the first
order, the fragmentation function of a quark is exactly
proportional to the square of its fractional charge, $e_i^2$.
This charge scaling is only slightly violated at small
$z$ for large $Q$ due to the gluonic contribution to the
QCD evolution as seen in Eq.~\ref{eq:evq}.

\section{ASSOCIATED DILEPTON PRODUCTION}

\subsection{Kinematical limits}

        In this paper, we are interested in the
dilepton production cross section integrated over
the transverse momentum. Hence,
we need only the dilepton fragmentation
functions integrated over $z$. As we have seen
in the previous Section, the fragmentation
functions diverge at $z=0$. One must therefore introduce
an infrared cutoff. Fortunately, for dilepton
production, the invariant mass $M$ provides
a natural cutoff.

Assuming that $Q^2$ and $q^2$ are the virtualities
of the parton before and after the emission of a
virtual photon with fractional momentum $z$, one can
verify that the relative transverse momentum of the
dilepton with respect to the original parton is
\begin{equation}
        k_T^2=z(1-z)\left[Q^2-\frac{M^2}{z}
                -\frac{q^2}{1-z}\right].
\end{equation}
Neglecting $q^2$ and requiring $k^2_T\ge0$,
we can see that $M^2$  provides a natural
kinematical cutoff for $z$,
\begin{equation}
        z\geq z_0\equiv M^2/Q^2. \label{eq:z0}
\end{equation}
In principle, one could take into account these
kinematical limits at every step of the radiation
processes, as done in Monte Carlo approaches \cite{KKEP,WEBB,PYTH,ODOR}.
Although not shown here, this can be done
analytically for the first order calculation of
the dilepton fragmentation functions.
One could also use the relative transverse momentum
$k_T^2$ as the argument in the running strong coupling
constant.  This is, however, beyond the scope of our simple
leading logarithm estimates in this paper.

        With the kinematical cutoff in Eq.~\ref{eq:z0}, we
can obtain the integrated dilepton fragmentation functions,
$D_{{\rm DL}/q_i,g}(M^2,Q^2)$, the probabilities for a quark
or gluon to produce a dilepton with mass $M$ within the
interval $dM^2$. The lowest and first order fragmentation
functions  can be obtained analytically by integrating
Eqs.~\ref{eq:dq03}, \ref{eq:dq11} and \ref{eq:dg11} over $z$,
\begin{eqnarray}
        D_{{\rm DL}/q_i}^{(0)}(M^2,Q^2) &=&
        e_i^2\left(\frac{\alpha}{2\pi}\right)^2\frac{2}{3M^2}
        \ln(\frac{M^2}{\Lambda^2})(e^{\kappa}-1) \nonumber\\
        &&\left\{
        2\ln(\frac{Q^2}{M^2})-\frac{3}{2}+2\frac{M^2}{Q^2}
        -\frac{1}{2}\frac{M^4}{Q^4}\right\},\label{dq0}\\
                                \nonumber\\
        D_{{\rm DL}/q_i}^{(1)}(M^2,Q^2) &=&
        e_i^2\left(\frac{\alpha}{2\pi}\right)^2\frac{2}{3M^2}
        \ln(\frac{M^2}{\Lambda^2})\frac{2}{\beta_0}
        (e^{\kappa}-1-\kappa)\frac{4}{3} \nonumber\\
        &&\biggl\{4g_2(\frac{M^2}{Q^2})-\frac{2\pi^2}{3}-
        \left(3-\frac{M^2}{Q^2}\right)
        \left(1-\frac{M^2}{Q^2}\right)\ln(1-\frac{M^2}{Q^2})
                                \nonumber\\
        &-& \frac{M^2}{Q^2}\left(2-\frac{1}{2}
        \frac{M^2}{Q^2}\right)\ln(\frac{M^2}{Q^2}) +
        \left(1-\frac{M^2}{Q^2}\right)\left(\frac{5}{2}-
        \frac{1}{2}\frac{M^2}{Q^2}\right)
        \biggr\},\label{dq1}\\
                \nonumber\\
        D_{{\rm DL}/g}^{(1)}(M^2,Q^2) &=&
        \left(\frac{\alpha}{2\pi}\right)^2\frac{2}{3M^2}
        \ln(\frac{M^2}{\Lambda^2})\frac{2}{\beta_0}
        (e^{\kappa}-1-\kappa)\frac{1}{2}\sum_i^{2n_f}e_{q_i}^2
                                \nonumber\\
        &&\biggl\{
        \left(\frac{4}{3}+2\frac{M^2}{Q^2}
                +\frac{M^4}{Q^4}\right)\ln(\frac{Q^2}{M^2})
                        \nonumber\\
        &-& \frac{1}{9}\left(1-\frac{M^2}{Q^2}\right)
        \left(22+13\frac{M^2}{Q^2}+4\frac{M^4}{Q^4}\right)
                        \biggr\}, \label{dg1}
\end{eqnarray}
where the function $g_2(x)$ is defined as
\begin{equation}
g_2(x)=\sum_{n=1}^{\infty}\frac{x^n}{n^2}.
\end{equation}

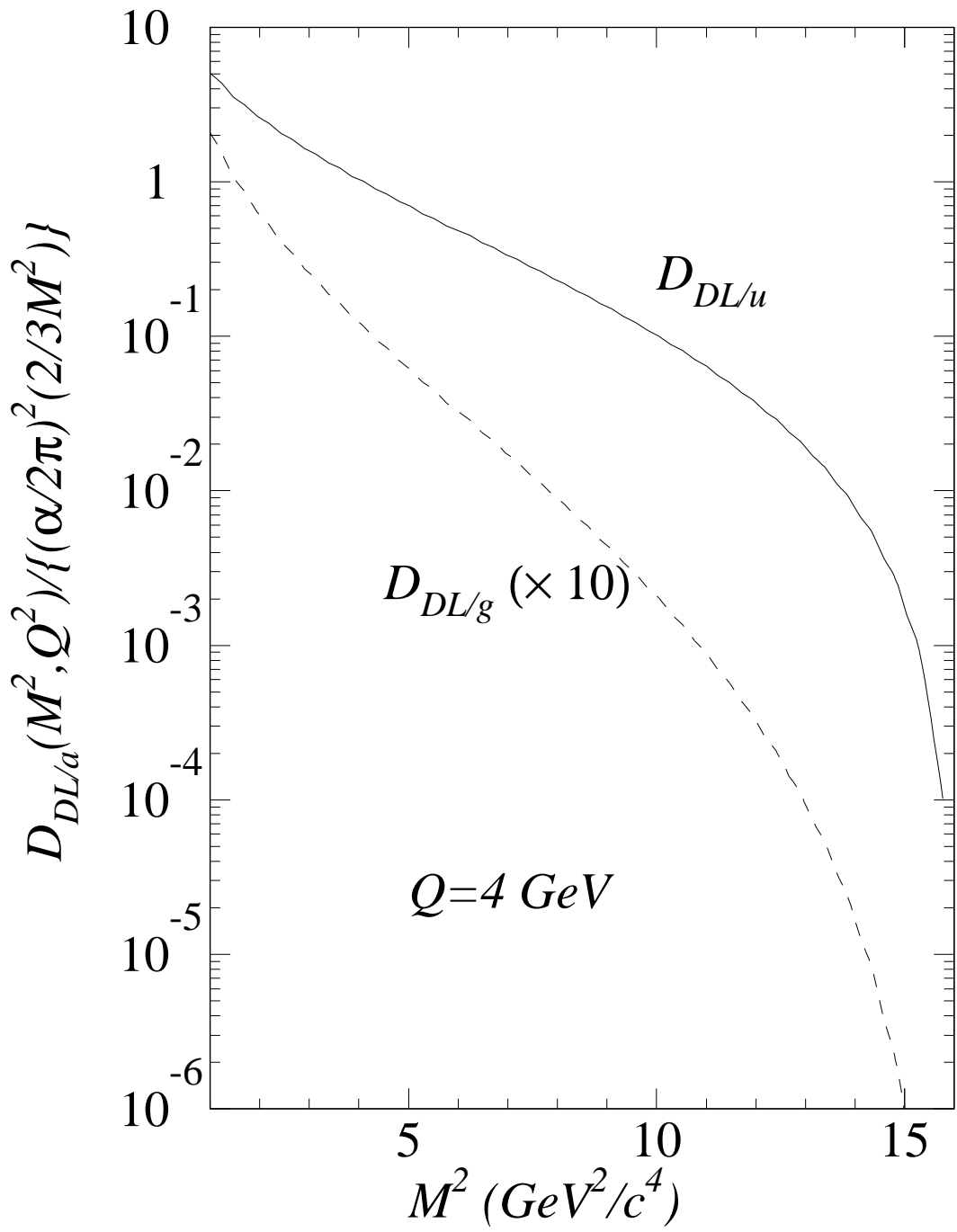
\begin{figure}
\caption{The $z$-integrated dilepton fragmentation functions
        $D_{{\rm DL}/a}(M^2,Q^2)$ for a $u$-quark (solid) and
        a gluon (dashed)
        as functions of $M^2$ at fixed $Q=4$ GeV. A factor
        $(\alpha/2\pi)^22/3M^2$ is divided out. The curve
        for gluon fragmentation is multiplied by 10.}
\label{fig3}
\end{figure}

        We plot in Fig.~\ref{fig3} the full
QCD-evolved results $D_{{\rm DL}/q_i,g}(M^2,Q^2)$
as functions of $M^2$ at fixed $Q$= 4 GeV.
The first order results $D_{{\rm DL}/q_i}^{(0)}(M^2,Q^2)
+D_{{\rm DL}/q_i}^{(1)}(M^2,Q^2)$ and
$D_{{\rm DL}/g}^{(1)}(M^2,Q^2)$ are very close
to the full QCD-evolved fragmentation functions
with only a few percent difference through the
whole $M^2$ range.
As we have seen in Fig.~\ref{fig2}, the full
QCD-evolved fragmentation functions are enhanced at small
$z$ while depleted at large $z$ as compared to the lowest
order calculations. For small values of $M^2/Q^2$, QCD evolution
is important, but the lower limit $z_0$ of the $z$-integration
is also small. Thus, the integrated full QCD
fragmentation functions are almost
the same as the first order results. At large values of
$M^2/Q^2$, the lower limit $z_0$ is large, but
the QCD corrections in any order are increasingly
smaller. Therefore, in the whole range of $M^2$, the first order
calculation of the $z$-integrated dilepton fragmentation
functions is a very good approximation.

\subsection{Dilepton production associated with minijets}

        In the following, we consider dilepton
production  associated with minijets. In particular,
we are interested in the differential rates of
dileptons with rapidity $Y=0$ as functions of the
invariant mass $M$. As a first approximation, we
can assume the dilepton to be produced collinearly
with the parent quark or gluon. Then
the differential cross section can be written down
in a straightforward manner by folding the
$z$-integrated dilepton  fragmentation functions
$D_{{\rm DL}/q_i}(M^2,Q^2)$ and $D_{{\rm DL}/g}(M^2,Q^2)$
with the $2\rightarrow2$ subprocesses of minijet
production.  We can also neglect the contribution
from the initial state dilepton radiation,
since the rapidities of these pairs are typically
large like those of the initially radiated
partons \cite{KEXW}.  One has to take into account
that the dilepton pair can be produced by either one
of the final state partons, and  connect this
to the correct normalization of the integrated
minijet cross section $\sigma_{\rm jet}$. In this way,
the basic formula for the associated production of
dileptons with $Y\sim 0$ from minijets in a $AA$
collision at impact parameter $b$ can be written as follows:
\begin{eqnarray}
        &&\frac{dN^{\rm DL/jet}_{AA}(b)}{dM^2 dY}
        = \frac{1}{2}\int d^2{r_{\perp}}\int_{p_0^2}^{s/4}
                        dp^2_Tdy_1 dy_2
        \sum_{{abcd=}\atop{q,\bar q,g}}
        x_1f_{a/A}(x_1,Q^2,r_{\perp})
        x_2f_{b/A}(x_2,Q^2,|{\bf b-r}_{\perp}|)  \nonumber\\
        &&\;\;\;\;\;\;\;\;\frac{d\sigma}{d\hat t}^{ab\rightarrow cd}
                \!\!\!\!\!\!\!\!\!\! \!\!(\hat s,\hat t,\hat u)
        \biggl\{D_{{\rm DL}/c}(M^2,Q_{\rm max}^2)\delta(Y-y_1) +
        D_{{\rm DL}/d}(M^2,Q_{\rm max}^2)\delta(Y-y_2)\biggr\},
                                \label{mini}
\end{eqnarray}
where the produced (anti)quarks and gluons (i.e. minijets)
have transverse momentum $p_0\le p_T\le \sqrt s/2$
and the kinematical range of rapidities
\begin{eqnarray}
        |y_1|\le \ln\bigl(\frac{\sqrt s}{2p_T}
                &+& \sqrt{\frac{s}{4p_T^2}-1}\bigr)\\
        -\ln\left(\frac{\sqrt s}{p_T}-{\rm e}^{-y_1}\right)\le
                &y_2& \le \ln\left(\frac{\sqrt s}{p_T}
                        -{\rm e}^{y_2}\right).
\end{eqnarray}
The momentum fractions of the initial state partons
are denoted by
\begin{equation}
        x_{1,2} = \frac {p_T}{\sqrt s}\left({\rm e}^{\pm y_1}
                +{\rm e}^{\pm y_2}\right),
\end{equation}
and the Mandelstam variables in the parton-parton
level for the massless partons by
\begin{eqnarray}
        \hat s &=& x_1x_2s=2p_T^2(1+\cosh(y_1-y_2)), \label{eq:shat}\\
        \hat t &=& -p_T^2(1+{\rm e}^{y_2-y_1}),\\
        \hat u &=& -p_T^2(1+{\rm e}^{y_1-y_2}). \label{eq:uhat}
\end{eqnarray}
The cross sections $d\sigma^{ab\rightarrow cd}/d\hat{t}
\sim {\cal O}(\alpha_s^2)$ for the various
partonic subprocesses can be found e.g. in Refs.~\cite{OWEN,SARC}.
The parton density of a nucleus by our
definition is
\begin{equation}
        f_{a/A}(x,Q^2,r_{\perp})=t_A(r_{\perp})R_{a/A}(x,Q^2,r_{\perp})
        f_{a/N}(x,Q^2), \label{eq:rtsh}
\end{equation}
where $t_A(r_{\perp})$ is the thickness function of the
nucleus which is normalized to $\int d^2r_{\perp} t_A(r_{\perp})=A$.
The parton distribution in a nucleon is $f_{a/N}(x,Q^2)$,
and the ratio $R_{a/A}(x,Q^2,r_{\perp})$ for the nuclear
modifications to the parton distributions is both scale and impact
parameter dependent \cite{HIJING,ESKO91}. In the following,
we will approximate the impact parameter dependent ratio
$R_{a/A}(x,Q^2,r_{\perp})$ by its averaged value,
\begin{equation}
        R_{a/A}(x,Q^2)\equiv \frac{1}{A}\int d^2r_{\perp}
                t_A(r_{\perp})R_{a/A}(x,Q^2,r_{\perp}),
\end{equation}
so that
\begin{equation}
        f_{a/A}(x,Q^2,r_{\perp})\approx t_A(r_{\perp})
                f_{a/\langle N\rangle}^A(x,Q^2),
\end{equation}
where the effective parton distributions per nucleon in a
nucleus is defined as
\begin{equation}
        f_{a/\langle N\rangle}^A(x,Q^2)\equiv
                R_{a/A}(x,Q^2)f_{a/N}(x,Q^2).
\end{equation}

In this paper we will use the  set 1 of  the Duke-Owens
parton distributions \cite{DO84} for $f_{a/N}(x,Q^2)$.
We use the scale dependent nuclear
modifications for $R_{a/A}(x,Q^2)$ as studied
in \cite{ESKO92}. Especially,  we assume that
at the lowest scale $Q=2$ GeV, gluons are shadowed by
the same amount as the structure function $F_2^A$
in deeply inelastic $\ell A$ scatterings. Note that
the normalization of Eq.~\ref{mini} can be checked
by setting the $M^2$-integrated fragmentation functions to unity
and integrating over $Y$; this will give us
$2\sigma_{\rm jet}(p_0,\sqrt s)$ as expected when
integrating over the inclusive $2\rightarrow2$
scattering cross section.

        As usually in the case of pQCD calculations,
there are uncertainties in choosing
the momentum scales both in the parton distributions
and the fragmentation functions. We will choose the
scale entering the parton distributions to be the
transverse momentum of the jets, $Q=p_T$, for
Duke-Owens parametrization set 1
with $\Lambda=0.2$ GeV \cite{OWEN}.
The scale $Q^2_{\rm max}$ in the dilepton fragmentation
functions represents the maximum virtuality of the final
state parton before any radiation. As we have emphasized
in this paper, the dilepton fragmentation functions at
large fixed mass do not have nonperturbative contributions.
Therefore, unlike the scale in the parton distributions,
$Q_{\rm max}$ in large mass dilepton fragmentation
functions is not correlated with the choice of
$\Lambda$. Examining the matrix elements
of $a+b\rightarrow a+b+\gamma^*$
processes, one can find out that the scale entering
the leading logarithm term is one of the Mandelstam
variables, $\hat s, -\hat t, -\hat u$, depending on
the channel of the specific process. However, in
Eq.~\ref{mini}, we convolute the fragmentation
functions with jet cross sections which
include different channels and their interference
terms. Therefore, $Q_{\rm max}$ in Eq.~\ref{mini}
is only an effective momentum scale. From
Eqs.~\ref{eq:shat}-\ref{eq:uhat}, we know at least that
$Q_{\rm max}\geq p_T$. We will discuss the sensitivity
to the choice of $Q_{\rm max}$ when we present the results
of our calculation.

Note also that for the minijet production the lower
limit $p_0$ of the integration over $p_T$ is a
parameter which determines the division between
calculable ``hard'' and model-dependent ``soft''
processes.  Most of the minijets are produced with
$p_T\sim p_0\sim$ few GeV/c,
and they are basically nonresolvable as distinct
$E_T$-clusters, even in hadronic collisions \cite{UA1}.
The phenomenological value of $p_0$ depends on the
model for $\sigma_{\rm soft}$ of soft
processes, the parton distribution functions and
the corresponding scale choice. Since these issues
can not be addressed within pQCD, the possible range
of values of $p_0$ has to be determined phenomenologically,
in connection with a model for the soft contribution
$\sigma_{\rm soft}$ to the  particle production in
$pp$ and $p\bar p$ collisions \cite{HIJING,KGBM,SJOS,WANG93}.
We will use here $p_0=2$ GeV/c, as suggested and
studied in detail in Ref.~\cite{WANG91}.  Although
already exactly calculated for inclusive jet
production \cite{EllisKS}, the ${\cal O}(\alpha_s^3)$
contributions to the lowest order parton cross sections
are simulated here by an overall
factor $K\sim 2$. Clearly, the parameter $p_0$ depends
also on the size of the next-to-leading order  terms.
We want to point out, however, that the cross section
for the associated dilepton
production in Eq.~\ref{mini} is much
less sensitive to the choice of $p_0$ than the minijet
cross section itself. For $Q_{\rm max}=p_T$, the dilepton
fragmentation functions vanish for $M\geq p_T$. Whenever
$M>p_0$, $M$ takes over as an effective cutoff in the
integration over $p_T$ in Eq.~\ref{mini}. Therefore,
the cross section for the associated dilepton production
does not depend on the exact choice of $p_0$ at large $M$.

The symmetrized formula of Eq.~\ref{mini} can be somewhat
simplified by considering all the possible pairs of partons
in the initial and final  states, ${\langle ab\rangle}$,
${\langle cd\rangle}$. By changing the integration
variables $y_{1,2}$ into $-y_{2,1}$ appropriately in the
other half of the expression, and by using the
$\hat t, \hat u$-symmetries of the subprocess cross sections,
a $Y\leftrightarrow -Y$ symmetric formula can be
written down.  Especially, at $Y=0$ we get:
\begin{eqnarray}
        \frac{dN^{\rm DL/jet}_{AA}(b)}{dM^2 dY}&&\bigg|_{Y=0}
        = 2T_{AA}(b)\int _{p^2_0}^{s/4}dp^2_T dy_2
        \sum_{{\langle ab\rangle}\atop{\langle cd\rangle}}
        \frac{1}{1+\delta_{ab}}\frac{1}{1+\delta_{cd}}
        x_1f_{a/\langle N\rangle}^A(x_1,Q^2)
                x_2f_{b/\langle N\rangle}^A(x_2,Q^2)
                                \nonumber\\
        &&\biggl\{ D_{{\rm DL}/c}(M^2,Q^2_{\rm max})
        \frac{d\sigma}{d\hat t}^{ab\rightarrow cd}
        \!\!\!\!\!\!\!\!\!\! \!\!(\hat s,\hat t,\hat u) +
        D_{{\rm DL}/d}(M^2,Q^2_{\rm max})
        \frac{d\sigma}{d\hat t}^{ab\rightarrow cd}
        \!\!\!\!\!\!\!\!\!\! \!\!(\hat s,\hat u,\hat t)
                \biggr \}\bigg|_{y_1=0}, \label{minidl}
\end{eqnarray}
where $T_{AA}(b)=\int d^2r_{\perp}t_A(r_{\perp})t_A(|{\bf b-r}_{\perp}|)$
is the nuclear overlap function of the two colliding nuclei.

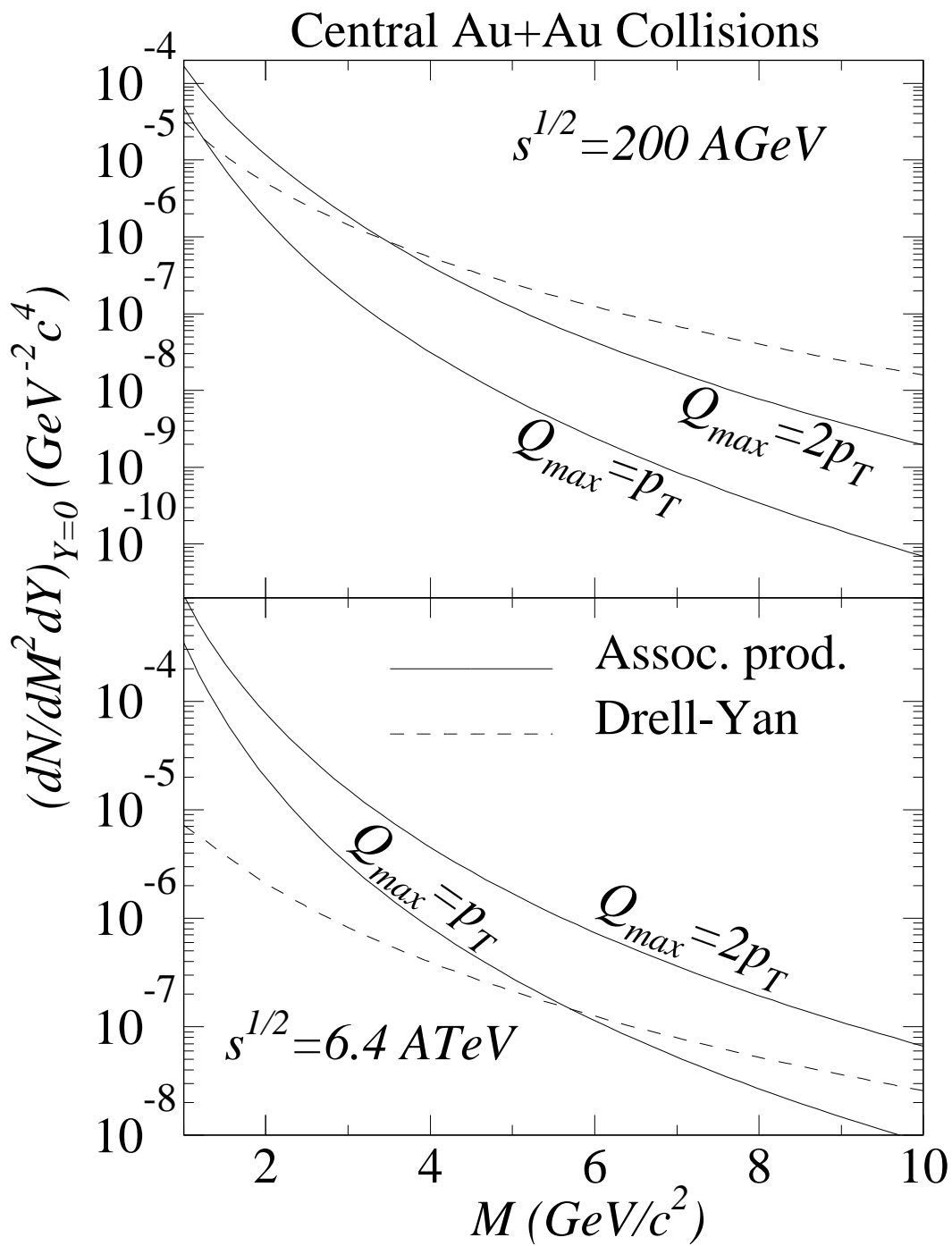
\begin{figure}
\caption{Mass spectra of minijet-associated (solid curves) and
        Drell-Yan (dashed) dileptons at $Y=0$ in
        central $Au+Au$ collisions at $\protect\sqrt{s}=200$
        and 6400 AGeV.
        The two solid curves
        correspond to two choices of the scale $Q_{\rm max}=p_T$
        and $2p_T$ in the dilepton fragmentation functions.
       Parton shadowing is included in the calculations.
       }
\label{fig4}
\end{figure}

As discussed in the previous Section, the first order
results in Eqs.~\ref{dq0}-\ref{dg1} are a good
approximation for the full $z$-integrated dilepton
fragmentation functions, which is the approximation
we shall adopt in what follows. The results from
Eq.~\ref{minidl} with nuclear modifications to the
parton distributions are shown  in Fig.~\ref{fig4} (solid curves)
for $\sqrt{s}=200$ AGeV and 6400 AGeV, respectively.
In the figure, we have compared the minijet associated
production of dileptons to the lowest order differential
cross section of the direct Drell-Yan process (dashed curves),
\begin{eqnarray}
        \frac{dN^{\rm DY}_{AA}(b)}{dM^2dY}\bigg|_{Y=0}
        &=& T_{AA}(b)\frac{4\pi\alpha^2}{9M^4}\sum_i e_{q_i}^2
        \biggl[x_1f_{q_i/\langle N\rangle}^A(x_1,M^2)
        x_2f_{\bar q_i/\langle N\rangle}^A(x_2,M^2) \nonumber\\
        &+& x_1f_{\bar q_i/\langle N\rangle}^A(x_1,M^2)
        x_2f_{ q_i/\langle N\rangle}^A(x_2,M^2)
                \biggr] , \label{drellyan}
\end{eqnarray}
where $x_{1,2}=M/\sqrt{s}$ at $Y=0$. We have chosen
the scale in the parton distributions as $Q=M$.
To simulate the first order pQCD contributions to
the DY cross section \cite{DYK1,FKWS}, we
multiply Eq.~\ref{drellyan} by an overall factor
$K_{\rm DY}\sim 2$. Note that since we have used the
Duke-Owens parton distributions, which extend only down
to $Q_0=2$ GeV, the results for direct Drell-Yan cannot
really be trusted much below $M=2$ GeV/c$^2$.

        To study the sensitivity of the minijet associated
dilepton production to the choice of the scale $Q_{\rm max}$
in the fragmentation functions, we plot in Fig.~\ref{fig4}
the results of Eq.~\ref{minidl} for both $Q_{\rm max}=p_T$
and $2p_T$. It is apparent that the results are relatively
sensitive to the choice of $Q_{\rm max}$. As we can understand
from Eqs.~\ref{dq0}-\ref{dg1}, the difference between the
two solid curves is due to the fact that the $z$-integrated
fragmentation functions are proportional to
$\ln^2(Q^2_{\rm max}/M^2)$. Due to the kinematical
restriction $M\leq Q_{\rm max}$, changing $Q_{\rm max}=2p_T$
to $p_T$ also effectively doubles the lower limit of
the integration over $p_T$ for fixed $M$ in Eq.~\ref{minidl}.
This is the reason why the two solid curves have different
slopes. As one of the main purposes of this paper,
Fig.~\ref{fig4} demonstrates how the relative
contribution of the dileptons associated with minijets
in the range $1\,{\buildrel < \over {_\sim}}\, M\,
{\buildrel < \over {_\sim}}\, 10$ GeV/c$^2$
changes with increasing energy as compared to the
direct Drell-Yan production. Even after taking into account
the uncertainties due to different choices of $Q_{\rm max}$,
it can be seen clearly that at RHIC energy, $\sqrt{s}=200$ AGeV,
dileptons from the bremsstrahlung of minijets are
comparable to the direct Drell-Yan at
$M\,{\buildrel < \over {_\sim}}\, 2$ GeV/c$^2$.
However, when going up to TeV energy range, dileptons
associated with minijets become more important, and
dominate the Drell-Yan at LHC energy, $\sqrt s=6400$ AGeV,
even up to masses $M\sim10$ GeV/c$^2$. Qualitatively, our results
are similar to the minijet-associated
photon production in Ref.~\cite{GUPTA}
where real photon fragmentation functions in the lowest
order are considered.

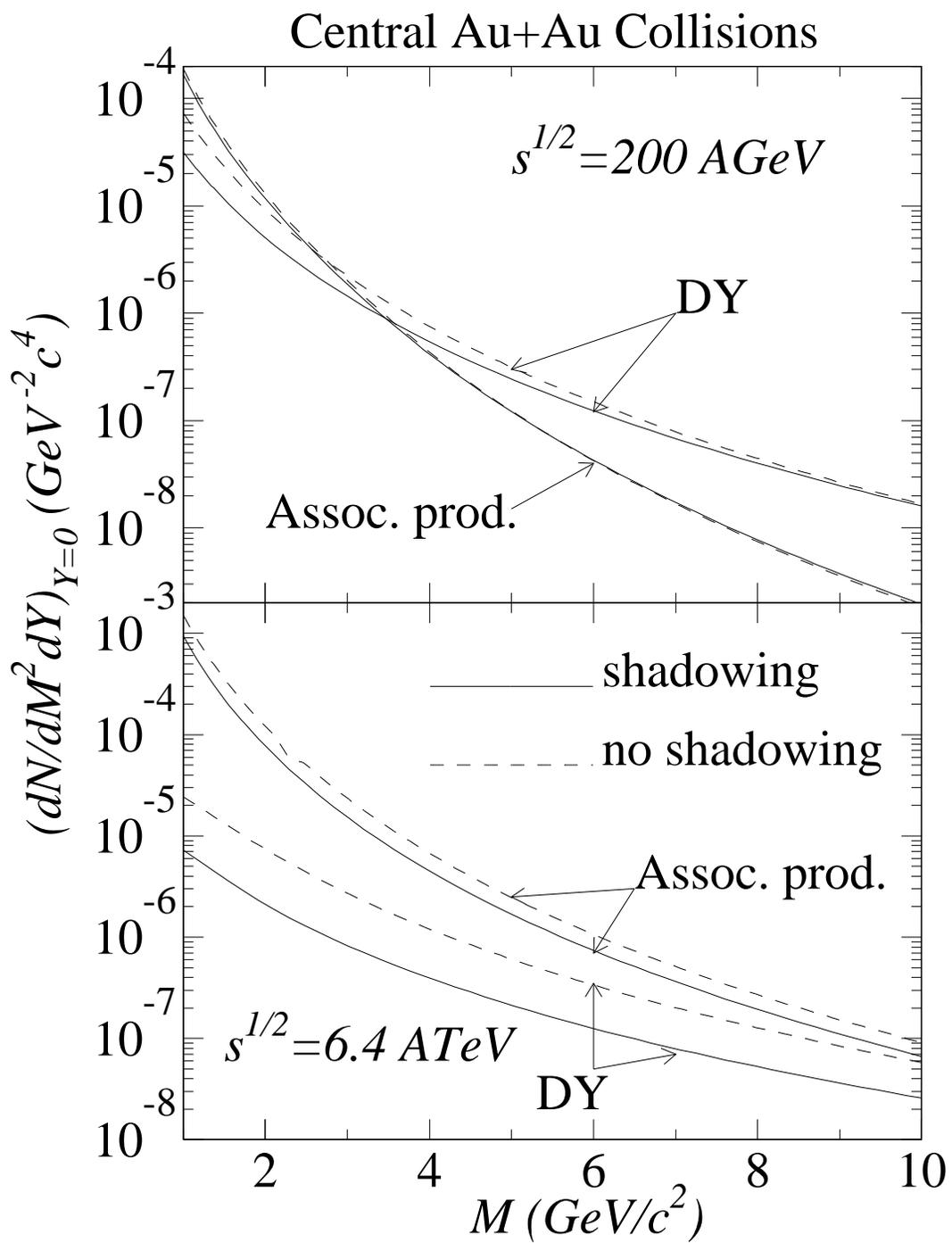
\begin{figure}
\caption{Mass spectra of the minijet-associated and Drell-Yan dileptons
        in central $Au+Au$ collisions at $\protect\sqrt{s}=200$
        and 6400 AGeV,
        with (solid) and without (dashed) parton shadowing.
        For the associated production, the scale in the dilepton
        fragmentation functions is chosen to be $Q_{\rm max}=2p_T$.
        }
\label{fig5}
\end{figure}

        To demonstrate the effects of parton shadowing
and antishadowing, we plot in Fig.~\ref{fig5} the
results calculated with (solid) and without (dashed curves)
nuclear modifications of the parton distribution
functions. We can see that nuclear shadowing
depletes the Drell-Yan dileptons relatively more than the
dileptons from the minijets. The basic reason for this
is that Drell-Yan pair production $dN^{DY}_{AA}$ in
Eq.~\ref{drellyan} as a function of $M=x_{1,2}\sqrt{s}$
probes the (anti)quark distributions directly,
at least in the lowest order. Furthermore, the antiquark
shadowing does not vary strongly with the  scale $M$,
as has been experimentally measured \cite{EMC,NMC}.
On the other hand, in the minijet-associated dilepton
production, we have to integrate the contribution over
the whole range of $x$.  In addition, we also have to
integrate over the scale $Q=p_T$. The gluon shadowing \cite{ESKO92}
we used here has stronger $Q$ dependence than the
(anti)quark. Therefore, the net effect of the nuclear modifications
of the parton distributions to the minijet-associated
dilepton production remains relatively small even at
TeV energy range.

\section{SUMMARY AND DISCUSSION}

        In this paper, we have studied minijet-associated
dilepton production in ultra-relativistic nuclear collisions.
We calculated both the first order approximation and the full
pQCD evolution of the dilepton fragmentation functions of
produced partons. The dilepton
pairs from the fragmentation of minijets are found to be
comparable to direct Drell-Yan at RHIC energy for small
invariant mass $M\sim$ 1--2 GeV/c$^2$. At LHC energy,
the associated dilepton production becomes dominant over
a relative large range of the invariant mass. These
dileptons plus the direct Drell-Yan pairs would
constitute part of the background to the dilepton
production from a QGP and its pre-equilibrium stage.
Other background includes dileptons from final hadronic
rescatterings \cite{KO,GALE} and the decay of
charmed hadrons \cite{GUPTA,VOGT}.

        It is also straightforward to calculate the
$p_T$ distribution of the associated dilepton pairs in
our fragmentation function approach. Since one has to
convolute the dilepton fragmentation functions in $z$
together with the $p_T$ distributions of the jets, we
expect the resultant $p_T$ spectrum of these dileptons
to be softer than the $p_T$ spectrum of the jets. Therefore,
the dileptons associated with minijets should have smaller $p_T$
relative to the direct Drell-Yan pairs which have a
high $p_T$ tail like that of the produced jets.
Since thermally produced dileptons in a QGP also have
relatively small $p_T$ as compared to Drell-Yan \cite{RUUS},
minijet-associated dileptons thus pose a more
intangible background.

        In calculating the dilepton fragmentation functions,
we have assumed leading logarithm approximation so that
we can include contributions from all orders in pQCD.
However, the higher order corrections are small and
the first order results are sufficient enough for our
estimates of the minijet-associated dilepton production.
The largest uncertainty in our calculation
is the choice of the momentum scale $Q_{\rm max}$
used in the dilepton fragmentation functions. Since the
correct scale in a matrix element calculation is
channel-dependent, we used only an effective scale choice in
the fragmentation functions to convolute with the minijet
cross sections. We evaluated the dilepton spectrum
for two choices of the scale, $Q_{\rm max}=p_T$, $2p_T$.
However, the results with $Q_{\rm max}=p_T$ should give us the
lower bound of the associated dilepton production.
Another notorious uncertainty of the $p_T$ cutoff $p_0$
in minijet-related problems is greatly reduced here
due to the kinematic restriction $M\leq Q_{\rm max}$.
For $Q_{\rm max}=p_T$, the $p_T$ cutoff is replaced
by $M$ whenever $M$ is larger than $p_0$.

        The abundance of dileptons associated with minijet
production at high energies is mainly due to the large
gluon-related minijet cross sections and the high initial
gluon densities inside the colliding nuclei. This should
have important implications for the dilepton production in
the  pre-equilibrium stage of the quark gluon plasma.
As pointed out recently \cite{BDMTW,GEIG93,ESKO93},
the parton system is not at all in chemical
equilibrium when initially produced in the earliest
stage of high energy nucleus-nucleus collisions.
Because of the small cross sections for (anti)quark
production, the initial parton system is dominated by
gluons and is quark deficient as compared to an
equilibrated QGP. Studies \cite{BDMTW,GEIG93,ESKO93}
also suggest that the parton system thus produced may not
be able to achieve chemical equilibrium before
hadronization. In this case, dilepton production
through $q\bar{q}$ annihilation should be severely
suppressed. On the contrary, dilepton production from
gluon fragmentation could become relatively important
for a gluon dominated system, since gluon-related cross
sections of small angle scatterings are about 9/4 larger
than the quark. Even though the dilepton fragmentation
function of a gluon is about one order of magnitude
smaller than a quark, a gluon density at least about
5 times higher than the quark could easily compensate
the small fragmentation function and make the
gluon associated dilepton production important.

\acknowledgments
We thank S.~Gupta and K.~Kajantie for helpful discussions.
KJE thanks Magnus Ehrnrooth foundation, Oskar \"{O}flund
foundation, and Suomen Kulttuurirahasto for partial financial support.
This work was supported by the Director, Office of Energy
Research, Division of Nuclear Physics of the Office of High
Energy and Nuclear Physics of the U.S. Department of Energy
under Contract No. DE-AC03-76SF00098.

\begin{references}
\bibitem[\dag]{address}Present address.
\bibitem{RUUS}P.~V.~Ruuskanen, Nucl. Phys. A {\bf 544},
                169c (1992).
\bibitem{LMCL}J.~I.~Kapusta, L.~McLerran and D.~K.~Srivastava,
        Phys. Lett. B {\bf 283}, 145 (1992).
\bibitem{KGJK}K.~Geiger and J.~I.~Kapusta, Phys. Rev. Lett.
        {\bf 70}, 1920 (1993).
\bibitem{SHXL}E.~Shuryak and L.~Xiong, Phys. Rev. Lett.
        {\bf 70}, 2241 (1993).
\bibitem{DY}S.~D.~Drell and T.-M. Yan, Phys. Rev. Lett.
                {\bf 25}, 316 (1970).
\bibitem{DYK1} G.~Altarelli, R.~K.~Ellis and G.~Martinelli,
        Nucl. Phys. B {\bf 157}, 461 (1979);
        J.~Kubar, M.~Le~Bellac, J.~L.~Meunier and G.~Plaut,
        Nucl. Phys. B {\bf 175}, 251 (1980).
\bibitem{FKWS}F.~Khalafi and W.~J.~Stirling, Z. Phys. C
        {\bf 18}, 315 (1983).
\bibitem{DYpT} Yu.~L.~Dokshitzer, D.~I.~D'yakonov and S.~I.~Troyan,
        Phys. Lett. B {\bf 78}, 290 (1978); Phys. Rep. {\bf 58}, 269 (1980);
        G.~Parisi and R.~Petronzio, Nucl. Phys. B {\bf 154}, 427 (1979);
        G.~Altarelli, R.~K.~Ellis, M.~Greco and G.~Martinelli,
        Nucl. Phys. B {\bf 246}, 12 (1984);
        G.~Altarelli, R.~K.~Ellis, and G.~Martinelli,
        Phys. Lett. B {\bf 151}, 457 (1985).
\bibitem{DYK2} A.~N.~Schellekens and W.~L.~van~Neerven,
        Phys. Rev. D {\bf 21}, 2619 (1980); {\em ibid.} {\bf 22}, 1623 (1980);
        T.~Matsuura, S.~C.~van~der~Marck and W.~L.~van~Neerven,
        Nucl. Phys. B {\bf 319}, 570 (1989);
        T.~Matsuura, R.~Hamberg and W.~L.~van~Neerven,
        {\em ibid.} {\bf 345}, 331 (1990).
\bibitem{KAJA} K.~Kajantie, P.~V.~Landshoff and J.~Lindfors,
        Phys. Rev. Lett. {\bf 59}, 2517 (1987); K.~J.~Eskola,
        K.~Kajantie and J.~Lindfors, Nucl. Phys. B {\bf 323}, 37 (1989).
\bibitem{HIJING} X.-N. Wang and M.~Gyulassy, Phys. Rev. D {\bf 44},
         3501 (1991); Phys. Rev. D {\bf 45}, 844 (1992).
\bibitem{KGBM}K. Geiger and B. M\"{u}ller, Nucl. Phys. B {\bf 369}, 600(1992);
        K. Geiger, Phys. Rev. D {\bf 47}, 133 (1993).
\bibitem{RANF}I. Kawrakow, H.-J. M\"{o}hring, and J. Ranft,
        Nucl. Phys. A {\bf 544}, 471c (1992).
\bibitem{OWEN}J.~F.~Owens, Rev. Mod. Phys. {\bf 59}, 465 (1987).
\bibitem{REYA}M.~Gl\"{u}ck, K.~Grassie, and E.~Reya, Phys. Rev.
        D {\bf 30}, 1447 (1984); {\em ibid.} {\bf 45}, 3986 (1992);
        {\em ibid.} {\bf 46}, 1973 (1992).
\bibitem{FIED}See e.g. R. D. Field, Applications of Perturbative
        QCD, {\sl Frontiers in Physics}, Vol. 77 (Addison-Wesley,
        1989).
\bibitem{AP}G.~Altarelli and G.~Parisi, Nucl. Phys. B {\bf 126},
        298 (1977).
\bibitem{DEWT}R.~J.~DeWitt, L.~M.~Jones, J.~D.~Sullivan,
        D.~E.~Williams, and H.~D.~Wyld, Jr., Phys. Rev. D
        {\bf 19}, 2046 (1979); {\bf 20}, 1751(E) (1979).
\bibitem{KKEP}K.~Kajantie and E.~Pietarinen, Phys. Lett. B
        {\bf 93}, 269 (1980).
\bibitem{WEBB}G.~Marchesini and B.~R.~Webber, Nucl. Phys.
        B {\bf 238}, 1 (1984).
\bibitem{PYTH} T.~Sj\"{o}strand, Comput. Phys. Commun. {\bf 39},
        347 (1986); T.~Sj\"{o}strand and M.~Bengtsson, {\em ibid.}
        {\bf 43}, 367 (1987).
\bibitem{ODOR}R. Odorico, Nucl. Phys. B {\bf 172}, 157 (1980).
\bibitem{KEXW}K.~J.~Eskola and X.-N. Wang, LBL preprint
        LBL-34156, 1993.
\bibitem{SARC}I.~Sarcevic, S.~D.~Ellis and P.~Carruthers,
        Phys. Rev. D {\bf 40}, 1446 (1989).
\bibitem{ESKO91}K.~J.~Eskola, Z. Phys. C {\bf 51}, 633 (1991).
\bibitem{DO84}D. W. Duke and  J. F. Owens, Phys. Rev. D
        {\bf 30}, 50 (1984).
\bibitem{ESKO92}K.~J.~Eskola, Nucl. Phys. B {\bf 400}, 240 (1993).
\bibitem{UA1}UA1 collaboration, C.~Albajar {\em et al.},
        Nucl. Phys. B {\bf 309}, 405 (1988).
\bibitem{SJOS}T.~Sj\"{o}strand and M.~van Zijl, Phys. Rev. D {\bf 36},
        2019 (1987).
\bibitem{WANG93} X.-N. Wang, Phys. Rev. D {\bf 46}, R1900 (1992);
                D {\bf 47}, 2754 (1993).
\bibitem{WANG91}X.-N.~Wang, Phys. Rev. D {\bf 43}, 104 (1991).
\bibitem{EllisKS} S.~D.~Ellis, Z.~Kunszt and D.~E.~Soper,
        Phys. Rev. Lett {\bf 64}, 2121 (1990);
        Z.~Kunszt and D.~E.~Soper, Phys. Rev. D {\bf 46}, 192 (1992).
\bibitem{GUPTA}S.~Gupta, Phys. Lett. B {\bf 248}, 453 (1990).
\bibitem{EMC}EM collaboration, M.~Arneodo {\em et al.},
        Nucl. Phys. B {\bf 333}, 1 (1990).
\bibitem{NMC}NM collaboration, P.~Amaudruz {\em et al.},
        Z. Phys. C {\bf 51}, 387 (1991).
\bibitem{KO}L.~Xiong, Z.~G.~Wu, C.~M.~Ko and J.~Q.~ Wu,
        Nucl. Phys. A {\bf 512}, 772 (1990).
\bibitem{GALE}C.~Gale and J.~I.~Kapusta, Phys. Rev. C {\bf 38},
        2659 (1988).
\bibitem{VOGT}R.~Vogt, B.~V.~Jacak, and P.~V.~Ruuskanen,
        Proceedings of Quark-Matter'93, Borl\"ange, Sweden, June
        20-24, 1993.
\bibitem{BDMTW}T. B. Bir\'{o}, E. van Doorn, B. M\"{u}ller, T. H. Thoma,
        and X. N. Wang, Duke University preprint DUKE-TH-93-46,
        to appear in Phy. Rev. C.
\bibitem{GEIG93}K.~Geiger and J.~I.~Kapusta, Phys. Rev. D
        {\bf 47}, 4905 (1993).
\bibitem{ESKO93}K.~J.~Eskola and M.~Gyulassy, Phys. Rev. C
        {\bf 47}, 2329 (1993).


\end{references}
\end{document}